\documentclass[12pt,preprint]{aastex}
\shorttitle{Unusual Asteroid 10537}
\shortauthors{Moskovitz et al. 2008}

\begin{document}

\title{A Spectroscopically Unique Main Belt Asteroid: 10537 (1991 RY16)}
\author{Nicholas A. Moskovitz}
\affil{Institute for Astronomy, 2680 Woodlawn Drive, Honolulu, HI 96822}
\email{nmosko@ifa.hawaii.edu}
\author{Samuel Lawrence}
\affil{School of Earth and Space Exploration, Arizona State University, PO Box 871404, Tempe, AZ 85287, sjlawren@asu.edu}
\author{Robert Jedicke, Mark Willman}
\affil{Institute for Astronomy, 2680 Woodlawn Drive, Honolulu, HI 96822, jedicke@ifa.hawaii.edu, willman@ifa.hawaii.edu}
\author{Nader Haghighipour}
\affil{Institute for Astronomy and NASA Astrobiology Institute, 2680 Woodlawn Drive, Honolulu, HI 96822, nader@ifa.hawaii.edu}
\author{Schelte J. Bus}
\affil{Institute for Astronomy, 640 North A'ohoku Place, Hilo, HI 96720, sjb@ifa.hawaii.edu}
\and
\author{Eric Gaidos}
\affil{Department of Geology and Geophysics and NASA Astrobiology Institute, University of Hawaii, POST 701, 1680 East-West Road, Honolulu, HI 96822, gaidos@hawaii.edu}

\begin{abstract}

We present visible and near-infrared reflectance spectra and interpreted surface mineralogy for asteroid 10537 (1991 RY16). The spectrum of this object is without precedent amongst the Main Belt asteroids. A unique absorption band centered at 0.63 microns could be attributed to one of several mineralogies. Pronounced 1- and 2-micron absorption bands suggest that the composition of 10537 is a mixture of pyroxenes and olivine and that it originated from a parent body that was partially or fully differentiated. The closest available analog is the large Main Belt asteroid 349 Dembowska but 10537 may be an isolated fragment from a completely eroded parent body.

\end{abstract}


\section{Introduction}

The spectra of asteroids have revealed many aspects of Solar System history. Differences between the spectra of S- and C-type asteroids are thought to represent a thermal gradient from the primordial solar nebula \citep{Gradie82}, the spectral slope of S-type asteroids gives clues to space weathering processes \citep{Nesvorny05}, and V-type asteroids trace the heating and melting of solid bodies during the epoch of planet formation \citep{Gaffey93b}. The discovery of new features in asteroid spectra provides insight to unstudied processes and mineralogies.  

In the early Solar System proto-planetary bodies were heated by the decay of short-lived radioactive isotopes (SLRs) such as $^{26}$Al and $^{60}$Fe \citep{Goswami05}. Temperatures in bodies that accreted quickly enough to incorporate high abundances of SLRs reached the solidus of silicates and partially melted, producing basaltic melt product \citep{Hevey06}. Isotopic analyses suggest that the iron and basaltic achondrite meteorites represent at least 60 differentiated parent bodies \citep{Chabot06,Yamaguchi01} and mineralogical analyses suggest that many asteroid families represent differentiated bodies \citep{Nathues05,Gaffey84,Mothe05,Sunshine04}, thus basaltic material should be common throughout the Main Belt. However, with the exception of Vesta and the dynamically associated Vestoids \citep{McCord70,Binzel93}, only three basaltic asteroids have been discovered \citep{Lazzaro00,Hammergren06,Binzel06,Roig08}. 

In the inner Main Belt ($a<2.5$ AU) the basaltic asteroid population is dominated by Vestoids, fragments that originated from the surface of Vesta in an impact event at least 3.5 Gyr ago \citep{Bottke05a}. The Yarkovsky effect and resonance scattering have made it difficult to map the dynamical boundaries of the Vestoid family \citep{Carruba05,Carruba07,Nesvorny08}, thus non-Vestoid basaltic asteroids  have yet to be unambiguously identified in the inner Main Belt.

The dynamics and in some cases the mineralogy of basaltic asteroids orbiting beyond the 3:1 mean motion resonance with Jupiter ($a>2.5$ AU) make them unlikely to have originated from the surface of Vesta \citep{Roig08,Hardersen04,Mich02}. Thus, the three basaltic asteroids in this region [1459 Magnya, 21238 (1995 WV7) and 40521 (1995 RL95)] likely represent independent cases of differentiation.

The paucity of non-Vestoid basaltic material in the Main Belt has motivated a number of recent searches \citep{Moskovitz08,Roig06,Hammergren06,Binzel06} that utilize the Sloan Digital Sky Survey Moving Object Catalog [SDSS MOC, \citet{Ivezic02}]  to identify taxonomically unclassified asteroids whose photometric colors indicate basaltic surface material. We present the optical and near-infrared (NIR) spectroscopic follow-up of one such asteroid, 10537 (1991 RY16). The orbital elements of 10537 ($a=2.85$ AU, $e=0.07$ and $i=7.25^\circ$) place it exterior to the 3:1 resonance, suggesting that it is not a fragment from Vesta.

\section{Observations}

We obtained low-resolution visible spectra of 10537 on 1 October 2006 with the Echellette Spectrograph and Imager (ESI) on the Keck II telescope \citep{Sheinis02}. Three 900-second exposures were obtained and an average of solar analog stars SA110-361, SA113-276 and SA93-101 were observed for calibration. Confirmation observations were performed on 18 January 2007 with the Supernova Integral Field Spectrograph (SNIFS) on the University of Hawaii 2.2 m telescope \citep{Lantz04}. Four 900-second exposures were obtained and solar analog star SA105-56 was observed for calibration. Details on the instrumental setup and reduction of ESI and SNIFS data are provided in \citet{Willman08} and \citet{Moskovitz08}. 

On 30 January 2008 NIR follow-up observations were performed with SpeX \citep{Rayner03} on NASA's Infrared Telescope Facility (IRTF). Forty-one 200-second exposures were obtained. The telescope was operated in a standard ABBA nod pattern and SpeX was configured in its low resolution (R=250) prism mode with a 0.8" slit for wavelength coverage from 0.8 - 2.5 $\mu$m. Solar analogs Hyades 64 and SA 102-1081 were observed for calibration and telluric correction. The IDL-based SpeXtool package \citep{Cushing04} was used for data reduction.

Figure \ref{fig.spec} shows that neither V-type asteroids Magnya and Vesta nor R-type asteroid 349 Dembowska (which we argue in \S4 is one possible parent body for 10537) are close spectral matches to 10537. Based on a chi-squared comparison of visible-NIR spectra, we do not find any V-type asteroids in the Main Belt or near-Earth populations that are as close a spectral match to 10537 as Dembowska. Relative to V-type asteroids, the $1~\mu$m band of 10537 is broader and its $2~\mu$m band is shallower and better matched by an R-type spectrum.

\section{Spectral Interpretation and Mineralogical Analysis}

\subsection{NIR Data}

As a first order approach to characterizing the spectrum of 10537 we performed a spectral band analysis \citep{Gaffey93a}. Without an asteroidal or meteoritic analog to the spectrum of 10537 or knowledge of its albedo, more sophisticated analyses based on Modified Guassian \citep[e.g.][]{Sunshine90} or Hapke mixing models \citep[e.g.][]{Lawrence07} are beyond the scope of this work.

With BI and BII referring to the 1- and 2-micron absorption bands respectively (Fig. \ref{fig.spec}) we find: BI central wavelength = $0.96 \pm0.01~ \mu$m, BII center = $1.91 \pm0.01~ \mu$m and the band area ratio (BAR) of BII  to BI = $1.22 \pm0.27$ (Fig. \ref{fig.gaffey}). The error bars on the band centers are equal to the width of the smoothing element used to fit the NIR data and the uncertainty in the BAR is based on 3-sigma error bars from the NIR spectrum. Following convention, we defined 2.5 $\mu$m as the red edge of BII. We have not corrected these parameters for temperature relative to Vesta because the noise in our spectrum is larger than typical temperature-induced changes. The BAR could be smaller depending on the actual shape of the 0.63 $\mu$m absorption feature, however we estimate a lower limit to the BAR of 0.9, a value which does not significantly affect the following discussion.

The location of 10537 in Fig. \ref{fig.gaffey} does not suggest a definitive mineralogy: both ordinary chondrites \citep[OCs,][]{Marchi05} and basaltic achondrites \citep[BAs,][]{Duffard05} are known to exist in the region of the band diagram occupied by 10537. Nevertheless, useful insight can be gained from Fig. \ref{fig.gaffey} in spite of the mineralogical ambiguity that is characteristic of this band analysis \citep{McCoy07}.

The location of 10537 above the olivine-orthopyroxene mixing line in Fig. \ref{fig.gaffey} suggests that its surface contains a mixture of low- and high-Ca pyroxene and olivine. The BI and BII centers imply the presence of Fe-rich low-Ca pyroxenes \citep{Adams74}. The pronounced 1- and 2-micron bands (Fig. \ref{fig.spec}) suggest negligible space weathering, a property that is characteristic of Vesta and the Vestoids \citep{Pieters00}. This ensemble of characteristics implies that 10537 may be achondritic and derived from a partially or fully differentiated parent body. 

\subsection{Visible Data}

10537 displays an unusually deep absorption band at $0.63~\mu$m whose mineralogical interpretation  is difficult because it is unprecedented amongst asteroids, rare in meteorites, and spans wavelengths that are populated with numerous solid state transitions. The eucrite meteorite Bouvante has a similar band around $0.65~\mu$m, which may be caused by ilmenite (FeTiO$_3$) or minor elements (such as Cr) in its pyroxene crystal structure \citep{Burbine01}. We suggest a number of other possibilities, however future studies will be necessary to properly constrain the mineralogical cause of this feature.

One possibility is Fe$^{3+}\rightarrow$ Fe$^{2+}$ charge transfer reactions between crystallographic sites in the pyroxene structure \citep{Mao72}. Ferric iron (Fe$^{3+}$) could have been inherited from the 10537 parent body, deposited on its surface by impacts with other bodies, or produced by shock-induced oxidation of Fe$^{2+}$ in pyroxenes \citep{Shes07}. The 0.63 $\mu$m feature could also be attributed to the presence of other transition metal cations. Cr$^{3+}$ is a likely candidate with other possibilities including Mn, V and Ti species \citep{Hiroi85,Hiroi88,Cloutis02}. Cr$^{3+}$ abundances greater than 1 oxide-weight-percent in terrestrial pyroxenes are sufficient to produce large absorption features at 0.63 $\mu$m \citep{Cloutis02}. Silicate Cr abundances are generally less than 1\% in the HED (Howardite-Eucrite-Diogenite) meteorites \citep{Mitt99} which are basaltic fragments thought to have originated from the surface of Vesta \citep{McCord70}. Thus it is not surprising that none of the Vestoids or HEDs display a prominent 0.63 $\mu$m band \citep{Binzel93,Bus02,Lazzaro04,Alvarez06,Mitt99}.

Another possible source for this feature is spinel \citep{Cloutis04}. In the NIR, spinel would increase the BAR without affecting the BI position, thus explaining why 10537 plots to the right of the Ol-Opx mixing line in Fig. \ref{fig.gaffey}. The presence of spinel, a product of igneous processes, would also suggest that 10537 is derived from a partially or fully differentiated parent body.

Other explanations for the 0.63 $\mu$m feature are less likely: a feature attributed to iron oxides in phyllosilicates on C-type asteroids \citep{Vilas93} occurs at slightly longer wavelengths ($\sim0.7~\mu$m), the faint iron oxide and spinel-group features observed in the spectra of S-types have band centers on either side of $0.63~\mu$m \citep{Hiroi96} and, although spin-forbidden Fe$^{2+}$ transitions were suggested by \citet{DiMartino95} to explain a similar (but not as deep) feature in the spectrum of V-type NEA 6611 (1993 VW), these transitions occur at shorter wavelengths than 0.63 $\mu$m \citep{Burns73}.

The prominence of this feature suggests a regolith with large grain sizes (e.g. a freshly exposed surface) or an unusually high abundance of the source mineral. It is unlikely that this feature is related to impact contamination which would have produced compositional heterogeneity across the surface, for which there is no evidence from our two visible spectra (Fig. \ref{fig.spec}).

\section{Discussion}

We have shown that 10537 is a spectroscopically unique asteroid whose composition may be suggestive of a partially or fully differentiated parent body. Its diameter is between 5 and 15 km, as determined from a plausible range of albedos (0.05-0.4) and an absolute magnitude of $H=12.9$, and therefore is unlikely to represent an intact differentiated body \citep{Hevey06} that has remained unfragmented since the time of planet formation \citep{Bottke05b}.

It is unlikely that 10537 is a scattered Vestoid because it orbits exterior to the 3:1 mean motion resonance \citep{Mich02,Roig08,Nesvorny08}. We numerically integrated the orbit of this asteroid and the four planets, Earth, Mars, Jupiter and Saturn using the N-body integration package MERCURY \citep{Chambers99}. Our simulations show that the orbit of 10537 is stable over 2 Gyr and presently is not in any mean-motion resonance with other planetary bodies.

\begin{figure}[b]
\plotone{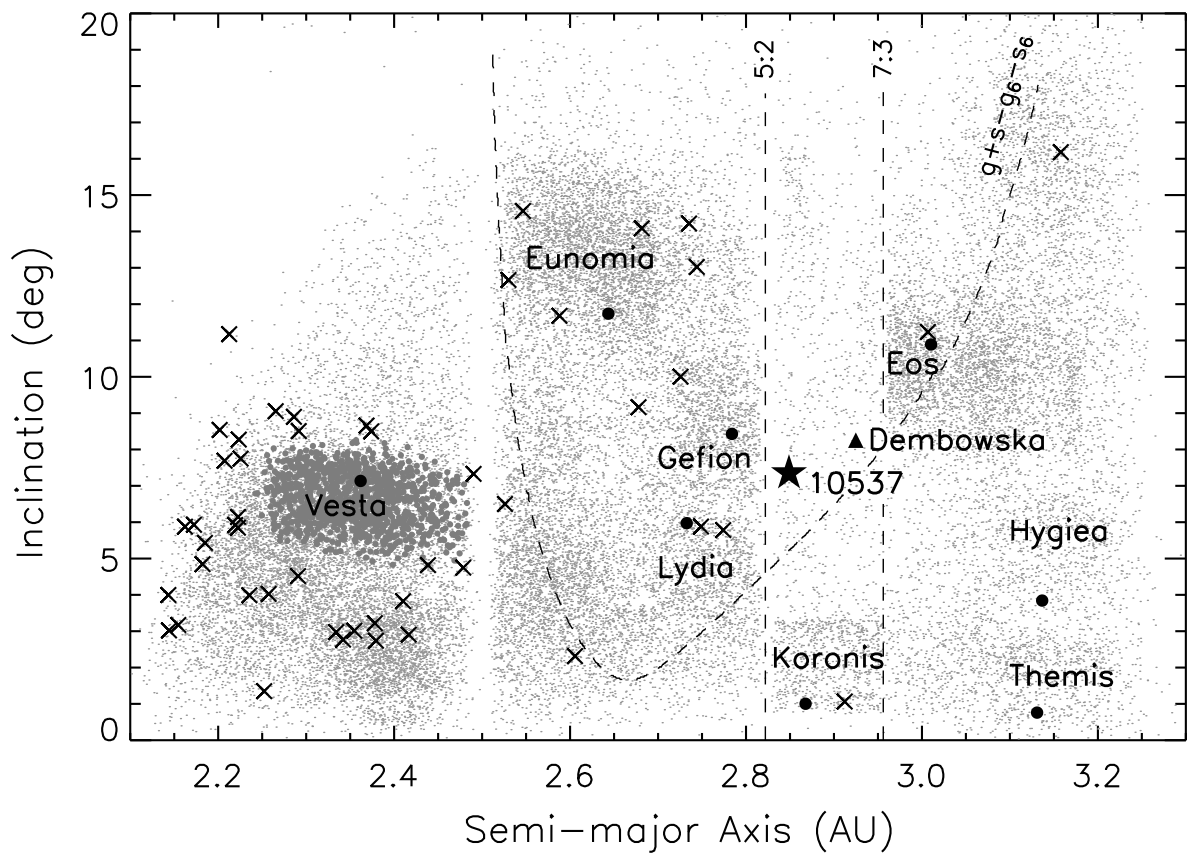}
\caption{Osculating semi-major axis versus inclination for SDSS MOC asteroids (small dots), our basaltic candidates ($\times$'s), Vestoids (filled grey circles), 10537 ($\star$) and Dembowska (triangle). The major families in the vicinity of 10537 are labeled and the parent body for each are plotted as filled black circles. 10537 shows no obvious correlation with any of these major families nor is it near any other of our basaltic candidates \citep{Moskovitz08}. The 5:2 and 7:3 mean motion resonances with Jupiter and a nonlinear secular resonance with Saturn are shown as dashed lines.}
\label{fig.orbits}
\end{figure}

10537 is not a member of any major asteroid family or near any other of our non-Vestoid basaltic candidates (Fig. \ref{fig.orbits}). Interactions with nearby non-linear secular resonances may have aided its orbital migration away from any parent body, particularly in inclination. Two such resonances are the $g+s-g_6-s_6$ and the $g+s-g_5-s_7$ \citep{Milani90, Milani92}. The former resonance for an eccentricity of 0.08 is shown in Fig. \ref{fig.orbits}. Accounting for the slightly larger eccentricity of 10537 would move this resonance to higher inclinations and closer to the orbit of 10537, thus increasing its relevance to the dynamical evolution of this body.

It is likely that the Yarkovsky effect also played a role in the orbital evolution of 10537. The maximum Yarkovsky drift in semi-major axis over 4.5 Gyr for a 5-km body (a lower limit for 10537) is $\sim0.1$ AU \citep{Bottke01}. This distance rules out the possibility that 10537 originated from the Eunomia or Eos families which may be derived from partially melted or differentiated parent bodies \citep{Nathues05,Mothe05}. Furthermore, 10537 would have had to pass through the 5:2 or 7:3 mean-motion resonance with Jupiter to reach its current orbit, a migration that would have been slower for a 5-km body \citep{Bottke01} than the dynamical lifetime of an asteroid in the 7:3 resonance \citep{Tsiganis03}.

One possible parent body for 10537 is the large ($\sim 140$ km) Main Belt asteroid 349 Dembowska. Dembowksa is one of only four known R-type asteroids in the Main Belt \citep{Bus02} and like 10537 has an uncertain surface mineralogy. \citet{Abell00} suggest that Dembowska has undergone some igneous processing and resembles an acapulcoite or a lodranite meteorite but has no known terrestrial analogue. \citet{Feierberg80} suggest that it is the cumulate mantle of a fully differentiated body. \citet{Gaffey93a} suggest it is the silicate residue from the partial melt of a partially differentiated body. It has also been suggested that Dembowska is a space weathered ordinary chondrite \citep{Hiroi01}.

\begin{figure}[b]
\plotone{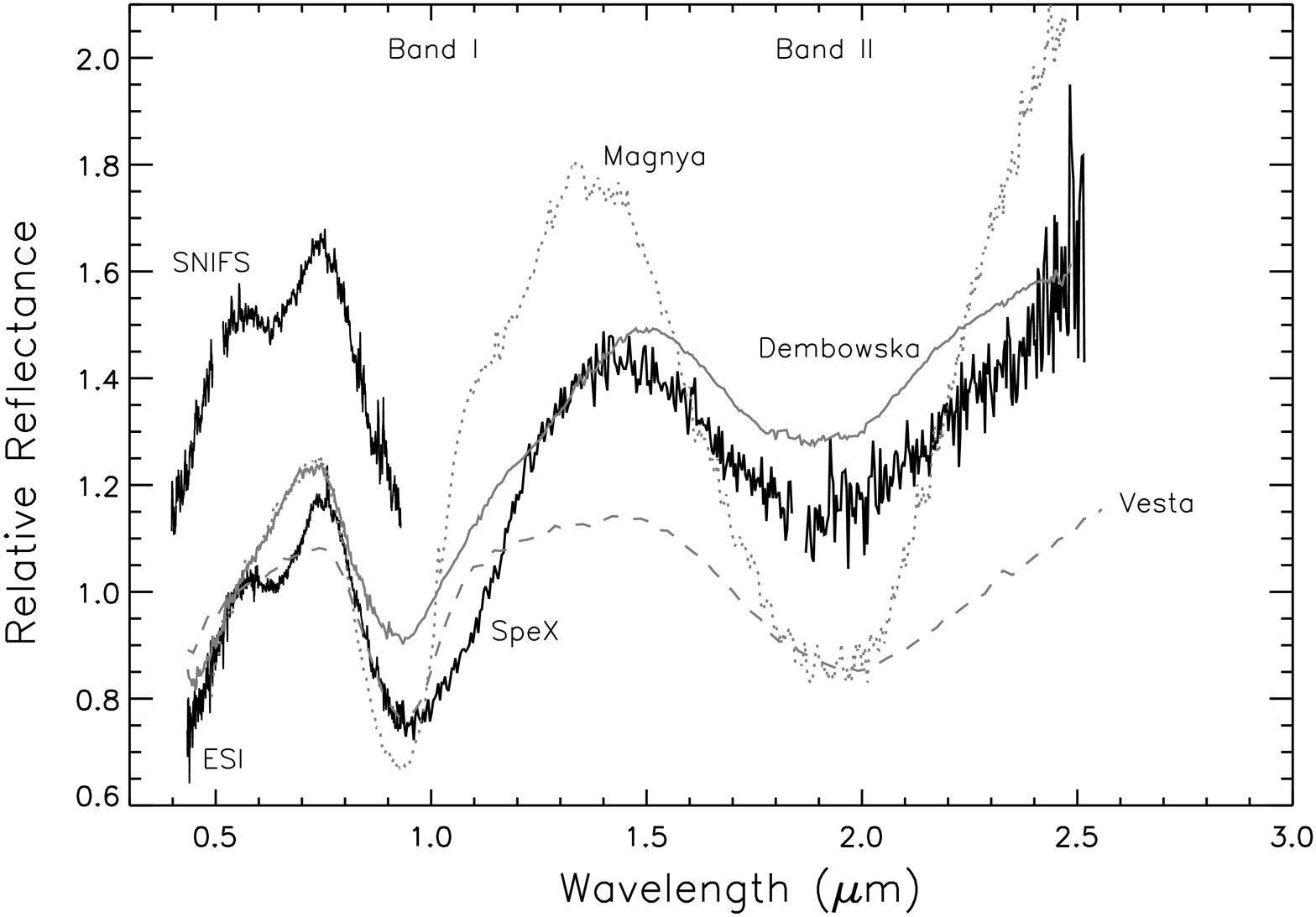}
\caption{ESI, SNIFS and SpeX spectra of asteroid 10537 (solid black lines). The transition from the ESI to SpeX data occurs at $0.96~\mu$m. The optical spectra have been normalized at 0.55 $\mu$m and the SNIFS data have been offset for clarity. Independent observations of 10537 are in agreement with these data \citep{Duffard08}. The NIR data have been normalized by a scaling factor that was calculated by minimizing the chi-squared statistic between the overlap of the ESI and SpeX data ($0.75 - 0.95~\mu$m). Spectra of V-type asteroids 1459 Magnya \citep{Hardersen04} and 4 Vesta \citep{Gaffey97} and R-type asteroid 349 Dembowska, all normalized at 0.55 $\mu$m, are shown for comparison.
}
\label{fig.spec}
\end{figure}

\begin{figure}[b]
\plotone{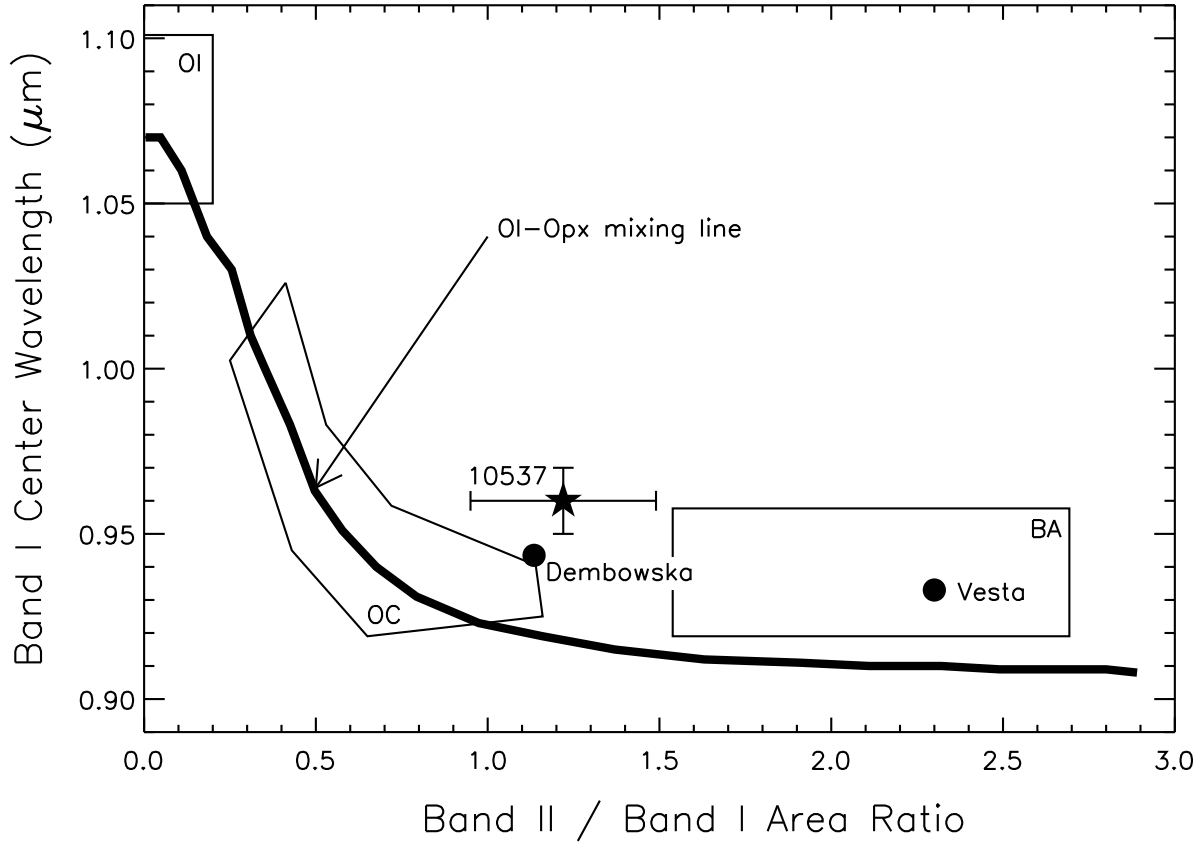}
\caption{Band diagram adapted from \citet{Gaffey93a}. The bold line indicates the olivine-orthopyroxene (Ol-Opx) mixing curve which traces a range of compositions from pure olivine (upper left) to 
pyroxene-dominated (lower right). The regions occupied by the ordinary chondrite (OC) and basaltic achrondite (BA) meteorites are outlined. Vestoids tend to fall within the BA region. 10537 is indicated by a star and Vesta and Dembowksa by filled circles \citep{Gaffey93a}. The errors on the Vesta and Dembowska values are comparable to the size of the symbols. Magnya plots off the figure to the right of Vesta with a BAR $\sim 4$ \citep{Hardersen04}.}
\label{fig.gaffey}
\end{figure}

Regardless of Dembowska's interpreted mineralogy, compositional heterogeneity across its surface \citep{Abell00} implies that a cratering event could have produced fragments like 10537. The spectral similarity between 10537 and Dembowska is consistent with this interpretation (Fig. \ref{fig.spec}):  the difference of $\sim10\%$ between these spectra is less that the spectral variation amongst the Vestoids. Until its cause is better understood, the 0.63 $\mu$m feature in the spectrum of 10537 does not argue against a genetic relation to Dembowska. Dembowksa's BAR of 1.135 and BI center of 0.9435 \citep{Gaffey93a} place it near 10537 in Fig. \ref{fig.gaffey}. From a purely dynamical stand point, a combination of ejection velocity \citep{Marchi04}, interactions with secular resonances and the influence of the Yarkovsky effect suggest that 10537 could have migrated to its current orbit from the surface of Dembowska. 

In spite of these similarities we note that Dembowska is merely the closest available analog to 10537 and may not be directly related in a petrogenetic sense. Under the standard paradigm of primordial clearing of the asteroid belt \citep{Bottke05b} most large bodies were disrupted, leaving behind representative fragments. This is a likely explanation for basaltic asteroid Magnya \citep{Mich02} and may apply to 10537 as well. 

Further observations coupled with more robust mineralogical analyses and detailed dynamical simulations will be necessary to better characterize 10537. A spectroscopic survey of asteroids in the dynamical space surrounding 10537 could reveal the presence of other similar bodies and establish or refute any connection to Dembowska. However, irrespective of whether these two asteroids are related, the spectral properties of 10537 indicate that there is more mineralogical diversity in the Main Belt than can be accounted for by traditional taxonomic definitions of asteroid families.

\acknowledgements Thanks to John Rayner for assistance with SpeX and to Ed Cloutis, Paul Hardersen and David Nesvorny for their helpful insights. N.M. is supported by a NASA GSRP fellowship. This work was partially supported under NSF grant AST04-07134 to R.J. Support by the NASA Astrobiology Institute under Cooperative Agreement NNA04CC08A for N.H. is acknowledged. The W.M. Keck Observatory is operated as a scientific partnership between Caltech, the University of California and NASA. The Observatory was made possible by the financial support of the W.M. Keck Foundation. The IRTF is operated by the University of Hawaii under a cooperative agreement with the NASA, Office of Space Science, Planetary Astronomy Program. We are deeply grateful to the native Hawaiian community who have allowed the use of Mauna Kea for an astronomical observatory.

\clearpage 

\clearpage

\clearpage

\begin{thebibliography}{}

\bibitem[Abell \& Gaffey(2000)]{Abell00} Abell, P.~A., \& Gaffey, M.~J. 2000, in LPSC Meeting 31 Proceedings, Abstract 1291

\bibitem[Adams(1974)]{Adams74} Adams, J.~B. 1974, J. Geophys. Res., 79, 4829

\bibitem[Alvarez-Candal et al.(2006)]{Alvarez06} Alvarez-Candal, A., Duffard, R., Lazzaro, D., \& Michtchenko, T. 2006, A\&A, 459, 969

\bibitem[Binzel \& Xu(1993)]{Binzel93} Binzel, R.~P., \& Xu, S. 1993, Science, 260, 186

\bibitem[Binzel, Masi, \& Foglia(2006)]{Binzel06} Binzel, R.~P., Masi, G., \& Foglia, S. 2006, in DPS Meeting 38 Proceedings, Abstract 71.06, Bulletin of the American Astronomical Society 38, 627

\bibitem[Bottke et al.(2005a)]{Bottke05a} Bottke, W.~F., Durda, D., Nesvorny, D., Jedicke, R., Morbidelli, A., Vokrouhlicky, D., \& Levison, H.~F. 2005a, Icarus, 175, 111

\bibitem[Bottke et al.(2005b)]{Bottke05b} Bottke, W.~F., Durda, D., Nesvorny, D., Jedicke, R., Morbidelli, A., Vokrouhlicky, D., \& Levison, H.~F. 2005b, Icarus, 179, 63

\bibitem[Bottke et al.(2001)]{Bottke01} Bottke, W.~F., Vokrouhlicky, D., Broz, M., Nesvorny, D., \& Morbidelli, A. 2001, Science, 294, 1693

\bibitem[Burbine et al.(2001)]{Burbine01} Burbine, T.~H., Buchanan, P.~C., Binzel, R.~P., Bus, S.~J., Hiroi, T., Hinrichs, J.~L., Meibom, A., \& McCoy, T.~J. 2001, Met. \& Planet. Sci., 36, 761

\bibitem[Burns et al.(1973)]{Burns73} Burns, R.~G., Vaughan, D.~J., Adu-Eid, M.~P., Witner, M., \& Morawsky, A. 1973, in Proc. Lunar Sci. Conf. 4th, 983

\bibitem[Bus \& Binzel(2002)]{Bus02} Bus, S.~J. \& Binzel, R.~P. 2002, Icarus, 158, 146

\bibitem[Carruba et al.(2005)]{Carruba05} Carruba, V., Michtchenko, T.~A., Roig, F., Ferraz-Mello, S., \& Nesvorny, D. 2005, A\&A, 441, 819

\bibitem[Carruba et al.(2007)]{Carruba07} Carruba, V., Roig, F., Michtchenko, T.~A., Ferraz-Mello, S., \& Nesvorny, D. 2007, A\&A 465, 315

\bibitem[Chabot \& Haack(2006)]{Chabot06} Chabot, N.~L., \& Haack, H. 2006, in Meteorites and the Early Solar System II, ed. D.~S. Lauretta, \& H.~Y. McSween Jr. (Tucson: Univ. Arizona Press), 747

\bibitem[Chambers(1999)]{Chambers99} Chambers, J.~E. 1999, MNRAS, 304, 793

\bibitem[Cloutis(2002)]{Cloutis02} Cloutis, E.~A. 2002, JGR, 107, E6, 5039

\bibitem[Cloutis et al.(2004)]{Cloutis04} Cloutis, E.~A., Sunshine, J.~M., \& Morris, R.V. 2004, Met. \& Planet. Sci., 39, 545

\bibitem[Cushing, Vacca \& Rayner(2004)]{Cushing04} Cushing, M.~C., Vacca, W.~D., \& Rayner, J.~T. 2004, PASP, 116, 362

\bibitem[Di Martino, Manara, \& Migliorini(1995)]{DiMartino95} Di Martino, M., Manara, A., \& Migliorini, F. 1995, A\&A, 302, 609

\bibitem[Duffard, Lazzaro, \& De Leon(2005)]{Duffard05} Duffard, R., Lazzaro, D., \& De Leon, J. 2005, Met. \& Planet. Sci., 40, 445

\bibitem[Duffard \& Roig(2008)]{Duffard08} Duffard, R., \& Roig, F. 2008, arXiv:0704.0230v2 [astro-ph]

\bibitem[Feierberg et al.(1980)]{Feierberg80} Feierberg, M.~A., Larson, H.~P., Fink, U., \& Smith, H.~A. 1980, GCA, 44, 513

\bibitem[Gaffey et al.(1993)]{Gaffey93a} Gaffey, M.~J., Bell, J.~F., Brown, R.~H., Burbine, T.~H., Piatek, J.~L., Reed, K.~L., \& Chaky, D.~A. 1993, Icarus, 106, 573

\bibitem[Gaffey, Burbine, \& Binzel(1993)]{Gaffey93b} Gaffey, M.~J., Burbine, T.~H., Binzel, R.~P. 1993, Meteoritics, 28, 161

\bibitem[Gaffey(1984)]{Gaffey84} Gaffey, M.~J. 1984, Icarus, 60, 83

\bibitem[Gaffey(1997)]{Gaffey97} Gaffey, M.~J. 1997, Icarus, 127, 130

\bibitem[Goswami et al.(2005)]{Goswami05} Goswami, J.~N., Marhas, K.~K., Chaussidon, M., Gounelle, M., \& Meyer, B.~S. 2005, in Proceedings of Chondrites and the Protoplanetary Disk, ASP Conference Series, ed. A.~N. Krot, E.~R.~D. Scott \& B. Reipurth, 341, 485

\bibitem[Gradie \& Tedesco(1982)]{Gradie82} Gradie, J., \& Tedesco, E. 1982, Science, 216, 1405

\bibitem[Hammergren, Gyuk, \& Puckett(2006)]{Hammergren06} Hammergren, M., Gyuk, G., \& Puckett, A. 2006, submitted to Icarus, preprint (astro-ph/0609420).

\bibitem[Hardersen, Gaffey, \& Abell(2004)]{Hardersen04} Hardersen, P.~S., Gaffey, M.~J., \& Abell, P.~A. 2004, Icarus, 167, 170

\bibitem[Hevey \& Sanders(2006)]{Hevey06} Hevey, P.~J. \& Sanders, I.~S. 2006, Met. \& Planet. Sci., 41, 95

\bibitem[Hiroi(1985)]{Hiroi85} Hiroi, T. 1985, M.Sc. thesis, (Tokyo: University of Tokyo)

\bibitem[Hiroi(1988)]{Hiroi88} Hiroi, T. 1988, Ph.D. thesis, (Tokyo: University of Tokyo)

\bibitem[Hiroi, Vilas, \& Sunshine(1996)]{Hiroi96} Hiroi, T., Vilas, F., \& Sunshine, J. 1996, Icarus, 119, 202

\bibitem[Hiroi \& Sasaki(2001)]{Hiroi01} Hiroi, T. \& Sasaki, S. 2001, Met. \& Planet. Sci., 36, 1587

\bibitem[Ivezic et al.(2002)]{Ivezic02} Ivezic, Z. et al. 2002, AJ 122, 2749

\bibitem[Lantz et al.(2004)]{Lantz04} Lantz, B. et al. 2004, in Proceedings of the SPIE, Optical Design and Engineering, ed. L. Mazuray, P.~J. Rogers, \& R. Wartmann, 5249, 155

\bibitem[Lawrence \& Lucey(2007)]{Lawrence07} Lawrence, S.~J., \& Lucey, P.~G. 2007, JGR, 112, E07005

\bibitem[Lazzaro et al.(2000)]{Lazzaro00} Lazzaro, D. et al. 2000, Science 288, 2033

\bibitem[Lazzaro et al.(2004)]{Lazzaro04} Lazzaro, D., Angeli, C.~A., Carvano, J.~M., Mothe-Diniz, T., Duffard, R., \& Florczak, M. 2004, Icarus, 172, 179

\bibitem[Mao \& Bell(1972)]{Mao72} Mao, H.~K., \& Bell, P.~M. 1972, Carnegie Inst. Washington Yearbook, 71, 524

\bibitem[Marchi, Lazzarin, \& Margin(2004)]{Marchi04} Marchi, S., Lazzarin, M., \& Margin, S. 2004, A\&A, 420, L5

\bibitem[Marchi et al.(2005)]{Marchi05} Marchi, S., Brunetto, R., Margin, S., Lazzarin, M., \& Gandolfi, D. 2005, A\&A, 443, 769

\bibitem[McCord, Adams, \& Johnson(1970)]{McCord70} McCord, T.~B., Adams, J.~B., \& Johnson, T.~V. 1970, Science, 168, 1445

\bibitem[McCoy et al.(2007)]{McCoy07} McCoy, T., Corrigan, C., Sunshine, J., Bus, S., \& Gale, A. 2007, in LPSC Meeting 38 Proceedings, Abstract 1338, 1631

\bibitem[Michtchenko et al.(2002)]{Mich02} Michtchenko, T.~A., Lazzaro, D., Ferraz-Mello, S., \& Roig, F. 2002, Icarus, 158, 343

\bibitem[Milani \& Knezevic(1990)]{Milani90} Milani, A. \& Knezevic, Z. 1990, CeMDA, 49, 347

\bibitem[Milani \& Knezevic(1992)]{Milani92} Milani, A. \& Knezevic, Z. 1992, Icarus, 98, 211

\bibitem[Mittlefehldt et al.(1999)]{Mitt99} Mittlefehldt, D., McCoy, T., Goodrich, C., \& Kracher, A. 1999, in Planetary Materials, Reviews in Mineralogy, ed. J. Papicke, 36, 4.1

\bibitem[Moskovitz et al.(2008)]{Moskovitz08} Moskovitz, N.~A., Jedicke, R., Gaidos, E., Willman, M., Nesvorny, D., Fevig, R., Ivezic, Z. 2008, submitted to Icarus

\bibitem[Mothe-Diniz \& Carvano(2005)]{Mothe05} Mothe-Diniz, T., \& Carvano, J.~M. 2005, A\&A, 442, 727

\bibitem[Nathues et al.(2005)]{Nathues05} Nathues, A., Mottola, S., Kaasalainen, M., \& Neukum, G. 2005, Icarus 175, 452

\bibitem[Nesvorny et al.(2008)]{Nesvorny08} Nesvorny, D., Roig, F., Gladman, B., Lazzaro, D., Carruba, V., \& Mothe-Diniz, T. 2008, Icarus, 193, 85

\bibitem[Nesvorny et al.(2005)]{Nesvorny05} Nesvorny, D., Jedicke, R., Whiteley, R.~J., \& Ivezic, Z. 2005, Icarus, 173, 132

\bibitem[Pieters et al.(2000)]{Pieters00} Pieters, P. et al. 2000, Met. \& Planet. Sci., 35, 1101 

\bibitem[Rayner et al.(2003)]{Rayner03} Rayner, J.~T., Toomey, D.~W., Onaka, P.~M., Denault, A.~J., Stahlberger, W.~E., Vacca, W.~D., Cushing, M.~C., \& Wang, S. 2003, PASP, 115, 362

\bibitem[Roig \& Gil-Hutton(2006)]{Roig06} Roig, F., \& Gil-Hutton, R. 2006, Icarus 183, 411

\bibitem[Roig et al.(2008)]{Roig08} Roig, F., Nesvorny, D., Gil-Hutton, R., \& Lazzaro, D. 2008, Icarus 194, 125

\bibitem[Sheinis et al.(2002)]{Sheinis02} Sheinis, A.~I., Bolte, M., Epps, H.~W., Kibrick, R.~I., Miller, J.~S., Radovan, M.~V., Bigelow, B.~C., \& Sutin, B.~M. 2002, PASP 114, 851

\bibitem[Shestopalov, McFadden, \& Golubeva(2007)]{Shes07} Shestopalov, D.~I., McFadden, L.~A., \& Golubeva, L.~F. 2007, Icarus, 187, 469

\bibitem[Sunshine, Pieters, \& Pratt(1990)]{Sunshine90} Sunshine, J.~M., Pieters, C.~M., \& Pratt, S.~F. 1990, JGR, 95, 6955

\bibitem[Sunshine et al.(2004)]{Sunshine04} Sunshine, J.~M., Bus, S.~J., McCoy, T.~J., Burbine, T.~H., Corrigan, C.~M., \& Binzel, R.~P. 2004, Met. \& Planet. Sci., 39, 1343

\bibitem[Tsiganis, Varvoglis, \& Morbidelli(2003)]{Tsiganis03} Tsiganis, K., Varvoglis, H., \& Morbidelli, A. 2003, Icarus, 166, 131

\bibitem[Vilas et al.(1993)]{Vilas93} Vilas, F., Larson, S.~M., Hatch, E.~C., and Jarvis, K.~S. 1993, Icarus, 105, 67

\bibitem[Willman et al.(2008)]{Willman08} Willman, M., Jedicke, R., Nesvorny, D., Moskovitz, N., Ivezic, Z., \& Fevig, R. 2008, Icarus, in press

\bibitem[Yamaguchi et al.(2001)]{Yamaguchi01} Yamaguchi, A., Misawa, K., Haramura, H., Kojima, H., Clayton, R.~N., Mayeda, T.~K., \& Ebihara, M. 2001, Meteoritics \& Planetary Science, 36, A228

\end{thebibliography}
\end{document}